\documentclass[twocolumn,tighten,times]{aastex701}

\usepackage{amsmath}
\usepackage{scalefnt}
\usepackage{fix-cm}

\newcommand{\anglest}{|\hat{\boldsymbol{j}}\cdot\hat{\boldsymbol{e}}_k|}

\newcommand{\eone}{\hat{\boldsymbol{e}}_1}
\newcommand{\etwo}{\hat{\boldsymbol{e}}_2}
\newcommand{\ethree}{\hat{\boldsymbol{e}}_3}

\shorttitle{Spin--Tidal Alignment in MaNGA}
\shortauthors{Moon \& Okumura}

\begin{document}

\title{Galaxy Spin Alignment with Tidal Fields in the SDSS-IV MaNGA Survey}

\author[0000-0001-7075-4156]{Jun-Sung Moon}
\affiliation{Institute of Astronomy and Astrophysics, Academia Sinica, 11F of Astronomy-Mathematics Building, No. 1, Sec. 4, Roosevelt Rd., Taipei 106319, Taiwan, R.O.C.}
\email{jsmoon.astro@gmail.com}

\correspondingauthor{jsmoon.astro@gmail.com}

\author[0000-0002-8942-9772]{Teppei Okumura} 
\affiliation{Institute of Astronomy and Astrophysics, Academia Sinica, 11F of Astronomy-Mathematics Building, No. 1, Sec. 4, Roosevelt Rd., Taipei 106319, Taiwan, R.O.C.}
\affiliation{Kavli IPMU (WPI), UTIAS, The University of Tokyo, Kashiwa, Chiba 277-8583, Japan}
\email{tokumura@asiaa.sinica.edu.tw}

\begin{abstract}
\hsize=180mm
\everypar={\leftskip=25pt}
\scalefont{0.97}
The tidal torque theory (TTT) predicts that galaxy spins are correlated with the surrounding tidal field, reflecting how angular momentum is acquired during structure formation. We present a new observational test of this prediction using the final data release of the Sloan Digital Sky Survey IV Mapping Nearby Galaxies at Apache Point Observatory integral field spectroscopy survey, which enables direct spin measurements from stellar and ionized gas kinematics for a sample of 6325 disk galaxies. We utilize the three-dimensional tidal field reconstructed from the galaxy distribution, providing a physically defined reference frame for the analysis. We find that massive galaxies tend to align their spins parallel to the intermediate axis of the tidal field, consistent with the prediction of the TTT, while also showing a tendency to align perpendicular to the major axis. In contrast, low-mass galaxies exhibit the opposite trend, with a transition mass of $M_* \sim 10^{10}$--$10^{10.5}\,M_\odot$. No significant alignment is detected with respect to the minor axis across all stellar masses. We further examine the dependence on morphology and environment, finding that S0 and early-type spiral galaxies exhibit stronger alignment signals than late-type spirals. The alignment trend becomes particularly pronounced in regions of high tidal anisotropy and high overdensity. A mutual information analysis identifies these environmental factors as the dominant drivers of the observed trends. Our results provide new empirical evidence for the connection between galaxy spins and the cosmic tidal field.
\end{abstract}



\scalefont{0.97}
\section{Introduction} \label{sec:intro}

Galaxy spins are not randomly oriented but exhibit preferred orientations in space. The origin of the galaxy spin alignment can be understood within the framework of the tidal torque theory (TTT), which suggests that a misalignment between a protohalo's inertia tensor and the surrounding tidal field generates a torque during the early stages of cosmic evolution (\citealt{1969ApJ...155..393P, 1970Ap......6..320D, 1984ApJ...286...38W}; see also \citealt{2002MNRAS.332..325P, 2002MNRAS.332..339P, 2020OJAp....3E...3N, 2021MNRAS.502.5528L, 2025MNRAS.543.2222L}). In the linear TTT, galaxy spins are expected to align preferentially with the intermediate axis of the local tidal tensor \citep{2000ApJ...532L...5L, 2009IJMPD..18..173S}. As structure formation proceeds, the direction of these spins can be altered by nonlinear vorticity in the cosmic flow \citep[e.g.,][]{2013ApJ...766L..15L, 2015MNRAS.446.2744L} and by mergers and tidal encounters between galaxies \citep[e.g.,][]{2007MNRAS.375..489H, 2009MNRAS.398.1742H, 2017MNRAS.465.2643C, 2021ApJ...909...34M}. Nevertheless, the spin directions remain intimately connected to the geometry of the cosmic web \citep[e.g.,][]{1996Natur.380..603B, 2008LNP...740..335V, 2024arXiv240716489K}, tracing how angular momentum builds up within the evolving large-scale structure \citep[e.g.,][]{2012MNRAS.427.3320C, 2015MNRAS.452.3369C}. Understanding how galaxy spins are coupled to their large-scale environments is therefore crucial for disentangling the competing physical processes that drive the acquisition and evolution of angular momentum in galaxies.

Cosmological $N$-body simulations show a mass-dependent trend in spin alignments, with a characteristic transition in the preferred spin orientation around a halo mass of $10^{12-13}\,M_\odot$ \citep{2007ApJ...655L...5A, 2007MNRAS.381...41H, 2009ApJ...706..747Z, 2012MNRAS.427.3320C, 2013MNRAS.428.2489L, 2013ApJ...762...72T, 2014MNRAS.440L..46A, 2014MNRAS.443.1090F, 2017MNRAS.468L.123W, 2018MNRAS.473.1562W, 2018MNRAS.481..414G, 2021MNRAS.503.2280G, 2020ApJ...902...22L, 2021MNRAS.502.5528L}. However, it remains uncertain whether a similar trend persists for observable galaxies, which probe only the inner regions of their host dark matter (DM) halos. Recent hydrodynamical simulations have predicted that stellar spins also exhibit clear alignment signals, albeit with a somewhat weaker mass-dependent transition than their DM counterparts \citep{2018MNRAS.481.4753C, 2018ApJ...866..138W, 2019MNRAS.487.1607G, 2020MNRAS.493..362K, 2021ApJ...922....6L, 2023ApJ...945...13M}.



Confirming the spin alignment observationally has long been challenging due to limited data availability. Early attempts inferred spin orientations indirectly from galaxy shapes \citep{2007ApJ...671.1248L, 2015ApJ...798...17Z, 2016MNRAS.457..695P}. However, the alignment between the shape and rotation axes holds only for galaxies that are predominantly rotation-supported. Defining the cosmic web adds further complications. Reconstructing the tidal field from galaxy distributions is nontrivial, and most observational studies have instead relied on filament orientations \citep{2010MNRAS.408..897J, 2013MNRAS.428.1827T, 2013ApJ...775L..42T, 2017A&A...599A..31H}, identified as density ridges in the galaxy distribution \citep[e.g.,][]{2011MNRAS.414..350S, 2015MNRAS.454.1140C, 2016A&C....16...17T}. These approaches are sensitive to the details of filament detection and to redshift-space distortions---particularly the finger-of-god effect---which introduces significant systematic uncertainties \citep[see, e.g.,][]{2013MNRAS.428.1827T, 2020MNRAS.491.2864W}.

The advent of integral-field spectroscopy (IFS) has transformed this field by allowing direct spin measurements from spatially resolved stellar kinematics. Using the Mapping Nearby Galaxies at Apache Point Observatory (MaNGA) survey \citep{2015ApJ...798....7B}, \citet{2019ApJ...876...52K} conducted the first measurement of spin alignments with nearby filaments but found no significant signal. Building on this work, \citet{2021MNRAS.504.4626K} analyzed 4633 MaNGA galaxies from the Sloan Digital Sky Survey (SDSS) Data Release 15 \citep{2019ApJS..240...23A} and measured three-dimensional (3D) spin--filament alignments, finding parallel alignments for spirals and perpendicular ones for S0s. \citet{2025ApJ...987L..30W} detected an antiparallel alignment signal between galaxy spins and the spins of their host filaments in MaNGA. Using the Sydney--AAO Multi-object Integral-field spectrograph (SAMI) Galaxy Survey \citep{2021MNRAS.505..991C}, \citet{2020MNRAS.491.2864W} detected a mass-dependent spin transition with respect to the filament spine. \citet{2022MNRAS.516.3569B} demonstrated spin alignment signals in about 3,000 SAMI galaxies, showing that the transition correlates most strongly with bulge mass. \citet{2023MNRAS.526.1613B} subsequently investigated how black hole activity relates to the spin alignment.

In this Letter, we present a new test of galaxy spin alignment with respect to the local tidal field using the latest IFS data from the MaNGA survey. This work represents the first study to examine spin alignment by combining (i) a reconstructed tidal field and (ii) kinematically measured spins. The use of the tidal field offers a distinct advantage for a physically well-motivated and consistently defined measure of the local gravitational deformation, applicable even in regions far from filament spines. Our analysis uses the final MaNGA data release \citep{2022ApJS..259...35A}, which contains nearly 10,000 galaxies, nearly doubling the sample size of previous IFS-based studies and enabling a statistically robust detection of alignment signals. These advances allow us to probe the intrinsic connection between galaxy spin and the underlying cosmic tidal field, offering a new window into the physics of galaxy formation within the cosmic web.


\section{Data and Methods} \label{sec:data}

\subsection{MaNGA Galaxies} \label{subsec:manga}

We utilize the IFS data from the complete release of the MaNGA survey, which is part of SDSS Data Release 17 \citep{2015ApJ...798....7B, 2022ApJS..259...35A}. All MaNGA data can be accessed via a Python-based tool called Marvin \citep{2019AJ....158...74C}. MaNGA employs a set of integral-field units, each consisting of between 19 and 127 fibers, to obtain spatially resolved spectra of nearby galaxies with a spectral resolution of $R\sim2000$ and a wavelength coverage of 3600--10400\,\AA. The raw spectra are flux-calibrated and combined into 3D data cubes through the MaNGA Data Reduction Pipeline (DRP; \citealt{2016AJ....152...83L, 2021AJ....161...52L}), and these data cubes are then processed by the MaNGA Data Analysis Pipeline (DAP; \citealt{2019AJ....158..160B, 2019AJ....158..231W}) to derive various astrophysical quantities, including stellar and gas kinematics, emission-line fluxes, and spectral indices. The DAP performs full-spectrum fitting on the binned spectra, providing the line-of-sight velocity fields for both the stellar and ionized gas components.


Our sample includes all galaxies observed in MaNGA, except those from four ancillary programs targeting nongalactic objects (\texttt{DEEP\_COMA}, \texttt{IC342}, \texttt{M31}, and \texttt{GLOBULAR\_CLUSTER}). We also exclude data flagged by the \texttt{DAPQUAL} bitmask as critical failures in either the DRP or DAP. Our parent sample consists of 10,058 unique galaxies, after removing duplicate entries in MaNGA. In this study, we include only disk galaxies (S0 and spiral types) classified in the MaNGA Visual Morphology Catalogue \citep{2022MNRAS.512.2222V}, resulting in a sample of 8199 galaxies.

We measure the position angles (PAs) of the kinematic major axis, $\phi_\mathrm{kin}$, of MaNGA galaxies using the PaFit package \citep{2006MNRAS.366..787K}, which determines the axis that maximizes the match between the velocity field and its bi-antisymmetry model. The central positions and redshifts of the galaxies are adopted from the MaNGA catalog, which for most galaxies is identical to the NASA--Sloan Atlas (NSA; \citealt{2011AJ....142...31B}). The MaNGA target selection strategy assigns 63\,\% of the main sample to reach 1.5 effective radii ($R_e$) and 37\,\% to reach $2.5\,R_e$ \citep{2015ApJ...798....7B}. For our analysis, we consistently use an elliptical aperture with a semi-major axis of $1.5\,R_e$ for each galaxy, with the aperture parameters defined by the NSA elliptical Petrosian photometry. We also exclude all bad pixels with inverse variance set to zero from the measurement. We assess the effect of the PA measurement uncertainties provided by PaFit through a Monte Carlo resampling test and find that their influence on the estimated spin alignment signal is negligible.

The spin axis is defined as perpendicular to the kinematic major axis. The vector direction is determined by identifying the approaching and receding sides from the inverse-variance-weighted averages of line-of-sight velocities on either side of the spin axis. Following the method of \citet{2007ApJ...671.1248L} and \citet{2011ApJ...732...99L}, we compute the 3D unit spin vector, $\hat{\boldsymbol{j}}$, in spherical coordinates as
\begin{gather} 
\hat{j}_r = \pm\,\cos{i},\label{eqn:spinr} \\ 
\hat{j}_\theta = -(1 - \cos^2{i})^{1/2} \cos{\vartheta},\label{eqn:spint} \\
\hat{j}_\phi = (1 - \cos^2{i})^{1/2} \sin{\vartheta},\label{eqn:spinp}
\end{gather}
where $\vartheta$ denotes the PA of the projected spin vector, measured eastward from north, ranging from $0$ to $2\pi$. The inclination angle $i$ is estimated under a circular disk approximation from the observed axial ratio $(b/a)$ and the intrinsic axial ratio $p$, and is given by
\begin{equation}
\cos^2{i} = \frac{(b/a)^2 - p^2}{1 - p^2},
\end{equation}
where we simply adopt $p = 0.158$ to be consistent with \citet{2021MNRAS.504.4626K}. 

Finally, the unit spin vector can be transformed into equatorial Cartesian coordinates as
\begin{gather}
\hat{j}_x = \hat{j}_r\,\cos{\delta} \cos{\alpha} + \hat{j}_\theta\,\sin{\delta} \cos{\alpha} - \hat{j}_\phi\,\sin{\alpha},\label{eqn:spinx} \\
\hat{j}_y = \hat{j}_r\,\cos{\delta} \sin{\alpha} + \hat{j}_\theta\,\sin{\delta} \sin{\alpha} + \hat{j}_\phi\,\cos{\alpha},\label{eqn:spiny} \\
\hat{j}_z = \hat{j}_r\,\sin{\delta} - \hat{j}_\theta\,\cos{\delta},\label{eqn:spinz}
\end{gather} 
with $\alpha$ and $\delta$ representing the R.A. and decl. of each galaxy, respectively. Due to the sign ambiguity of the radial component $\hat{j}_r$, the 3D spin vector cannot be uniquely determined without additional assumptions, resulting in a twofold degeneracy \citep{2011ApJ...732...99L, 2021NatAs...5..283M}. While many previous studies arbitrarily selected one of the two vectors \citep[e.g.,][]{2021MNRAS.504.4626K, 2022MNRAS.516.3569B}, we account for both in our analysis, which differ only by the sign of $\hat{j}_r$. We note that our results are not sensitive to either the sign of $\hat{j}_r$ or the value of $p$. 

\subsection{Reconstructed Tidal Field} \label{subsec:tweb}

We obtain the tidal field around the MaNGA galaxies from the Galaxy Environment for MaNGA Value Added Catalog (GEMA-VAC; \citealt{2015A&A...578A.110A, 2022ApJS..259...35A}). The GEMA-VAC provides a characterization of large-scale structures based on the Exploring the Local Universe with the reConstructed Initial Density field (ELUCID) project \citep{2016ApJ...831..164W}. The present-day density field within the SDSS volume is reconstructed from the distribution of galaxy groups using the halo-domain method \citep{2009MNRAS.394..398W, 2013ApJ...772...63W}. The survey volume is partitioned into halo domains, and the mass distribution in each domain is modeled with $N$-body simulations. Redshift-space distortions are iteratively corrected using the line-of-sight component of the reconstructed linear velocity field.

The tidal field ($T_{ij}$) is defined as the Hessian of the gravitational potential field ($\phi$):
\begin{equation}
T_{ij} = \frac{\partial^2 \phi}{\partial x_i \partial x_j},
\end{equation}
where $\phi$ is related to the reconstructed density field ($\rho$) through the Poisson equation:
\begin{equation}
\nabla^2 \phi = 4\pi G\rho.
\end{equation}
The density field is smoothed with a Gaussian kernel with a comoving scale of 2\,$h^{-1}$Mpc. The three eigenvalues of $T_{ij}$, $\lambda_1 \geq \lambda_2 \geq \lambda_3$, are obtained by diagonalizing $T_{ij}$ at the position of each galaxy. The corresponding eigenvectors are denoted by the major ($\hat{\boldsymbol{e}}_1$), intermediate ($\hat{\boldsymbol{e}}_2$), and minor ($\hat{\boldsymbol{e}}_3$) principal axes of $T_{ij}$, respectively. The major principal axis, $\hat{\boldsymbol{e}}_1$, corresponds to the direction of maximum matter compression. The eigenvalues can be used to classify the large-scale environment based on the number of positive eigenvalues: three positive eigenvalues indicate a cluster environment, two indicate a filament, one indicates a sheet, and none indicates a void. In our sample, a total of 6376 disk galaxies have corresponding tidal field data available for analysis. After excluding the small number lacking stellar mass or reliable PA, 6325 disk galaxies constitute the final analysis sample used throughout this work.

\subsection{Alignment Angles} \label{subsec:angle}

The spin--tidal alignment is quantified as the cosine of the angle, $\anglest$ ($k = 1, 2, 3$), where $\hat{\boldsymbol{j}}$ and $\hat{\boldsymbol{e}}_k$ represent the unit spin vector of each galaxy and the three principal axes of the local tidal tensor, respectively. In three dimensions, a random orientation yields an average of $\anglest = 0.5$, while parallel and perpendicular orientations correspond to $\anglest = 1$ and $0$, respectively. Since the line-of-sight component of spin vectors cannot be uniquely determined in this study, each galaxy has two possible 3D spin orientations. In the edge-on view ($i = \pi/2$), the two vectors appear identical, while in the face-on view ($i = 0$), they appear to point in opposite directions. To account for this ambiguity, we take the average of $\anglest$ from the two possible spin vectors to obtain a single alignment metric for each galaxy. This approach retains as much orientation information as possible while ensuring that the two spin vectors from the same galaxy are not treated as independent measurements. We also confirm that fixing the sign of $\hat{j}_r$ to either $+$ or $-$ leads to qualitatively identical alignment trends (see Appendix \ref{append:sign}), which is consistent with the findings of previous studies \citep[e.g.,][] {2021MNRAS.504.4626K, 2025ApJ...987L..30W}.

The detection significance of the observed alignment is estimated from the 1000 random permutation samples in which the sky positions ($\alpha$ and $\delta$) of galaxies are randomly shuffled within the sample. For each random permutation, the 3D spin vectors are recalculated using the new sky coordinates through Equations (\ref{eqn:spinr})--(\ref{eqn:spinz}). This approach provides the null hypothesis of random mutual orientation between galaxy spins and tidal fields, although it inevitably disrupts possible selection effects tied to galaxy sky positions and may therefore under- or overestimate the significance. We find, however, that alternative schemes (e.g., only randomizing $\phi_\mathrm{kin}$) do not alter the main conclusions of this work.


\section{Spin--Tidal Alignment} \label{sec:align}

\begin{figure*}[t]
\epsscale{1.15}
\plotone{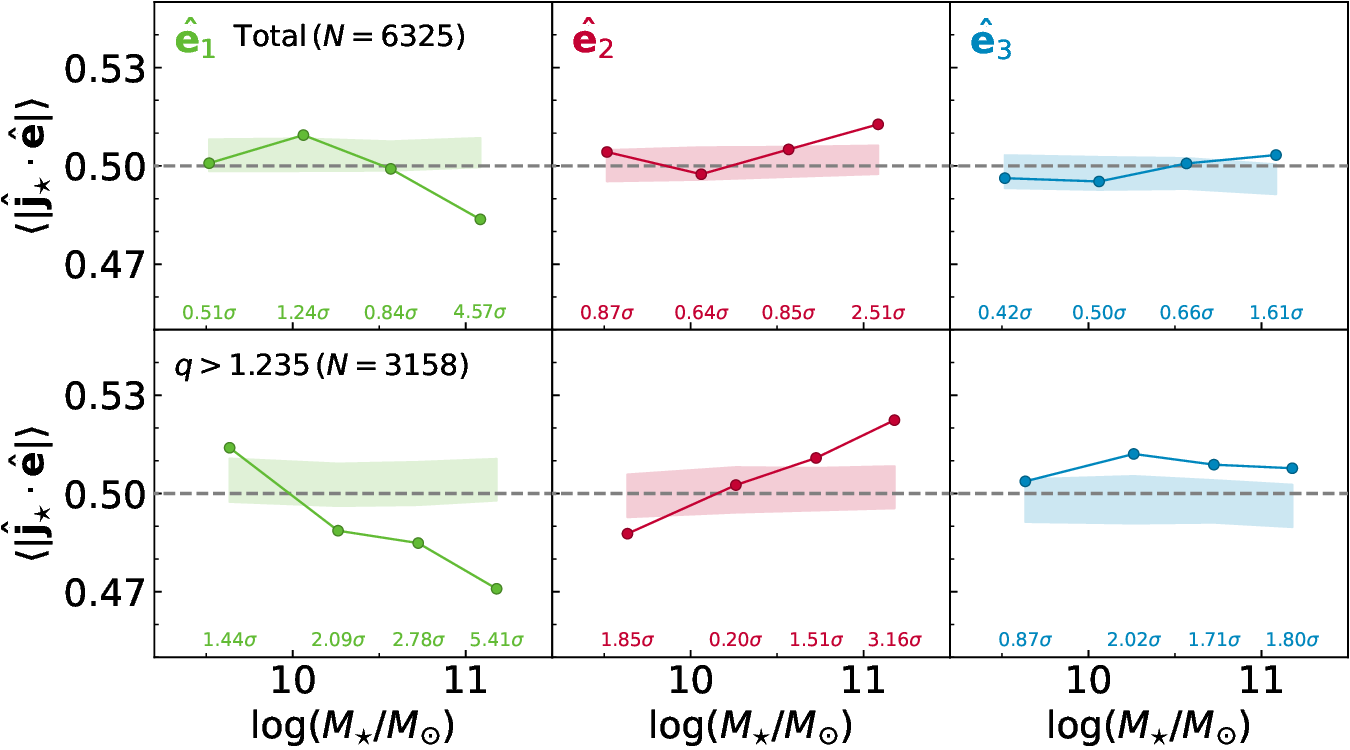}
\caption{Top: mean absolute values of the cosines of the angles between stellar spin vectors and the three principal axes of the local tidal tensor as a function of stellar mass. From left to right, the panels correspond to the major (green), intermediate (red), and minor (blue) principal axes. In each panel, the sample is divided into bins of equal size. The shaded regions indicate the 1$\sigma$ intervals from 1000 randomly shuffled samples, and the dashed lines mark the expected value for a random orientation. The detection significance for each bin is indicated below each point. Bottom: same as the top row, but for galaxies in regions of strong tidal anisotropy ($q > 1.235$). \label{fig:alignall}}
\end{figure*}

In this section, we present the main results of our analysis. Specifically, we examine how the MaNGA galaxies align their spins with respect to the principal axes of the local tidal field. The top row of Figure \ref{fig:alignall} shows the mean cosines of the angles between the stellar spin vectors and the three principal axes as a function of stellar mass. The corresponding results for the gas spin vectors are presented in Appendix \ref{append:gas} and Figure \ref{fig:aligngas}, and a consistent alignment trend is observed between the stellar and gas spins. The alignment signals reveal a clear mass-dependent trend. Notably, for galaxies more massive than $10^{10.5}\,M_\odot$, the spin vectors tend to be perpendicular to the major axis ($\eone$), where the detection significance reaches $4.6\sigma$. Additionally, massive galaxies exhibit a mild parallel alignment signal with the intermediate axis ($\etwo$), with a detection significance of $2.5\sigma$.

The bottom row of Figure \ref{fig:alignall} shows the spin--tidal alignments for galaxies residing in regions where the tidal anisotropy exceeds the sample median ($q_\mathrm{med} = 1.235$). The tidal anisotropy, $q$, quantifies the degree of anisotropy in the local tidal field \citep{2018MNRAS.476.3631P} and is defined as
\begin{equation}
q = \sqrt{\frac{(\lambda_1-\lambda_2)^2+(\lambda_1-\lambda_3)^2+(\lambda_2-\lambda_3)^2}{2}},
\end{equation}
where $\lambda_1$, $\lambda_2$, and $\lambda_3$ are the three eigenvalues of the local tidal tensor. Galaxies in regions of strong tidal anisotropy exhibit a distinct mass-dependent alignment, with a transition in the preferred alignment direction between the low- and high-mass regimes. Massive galaxies ($M_* \geq 10^{10.5}\,M_\odot$) align their spins perpendicular to the major axis ($\eone$) and parallel to the intermediate axis ($\etwo$), consistent with the trend seen in Figure \ref{fig:alignall}. The detection significance is substantially higher than that in Figure~\ref{fig:alignall}, reaching $5.4\sigma$ for the alignment with $\eone$ and $3.2\sigma$ for that with $\etwo$. In contrast, low-mass galaxies ($M_* < 10^{10}\,M_\odot$) show the opposite tendency, with their spins preferentially aligned parallel with $\eone$ and perpendicular to $\etwo$. With respect to the minor axis ($\ethree$), galaxies more massive than $10^{10}\,M_\odot$ show a weak alignment signal without any clear trend across this mass range.

The stronger alignment observed in high-$q$ regions is likely due to coherent gravitational influences from highly anisotropic tidal fields. In addition, the reconstructed tidal field tends to be more reliable where $q$ is high, because its principal axes become ambiguous when the three eigenvalues of the tidal tensor all have similar magnitudes. This ambiguity could contribute to the lack of a clear alignment signal in such regions. Moreover, the strength of the spin--tidal alignment appears to depend strongly on other environmental factors, as will be shown later, suggesting that the enhanced alignment in high-$q$ regions may arise from environmental effects.

Our analysis, combining a large IFS survey with a directly reconstructed tidal field, provides a more statistically robust detection of the spin--tidal alignment than previous observational studies. The results reveal a mass-dependent trend broadly consistent with recent state-of-the-art hydrodynamic simulations, which have reported an alignment between stellar spins and the cosmic web but only a weak signal for low-mass galaxies \citep{2018MNRAS.481.4753C, 2019MNRAS.487.1607G, 2020MNRAS.493..362K, 2021ApJ...922....6L, 2023ApJ...945...13M}. We find a stronger alignment for massive galaxies, with a higher detection significance than in lower-mass systems. Nevertheless, we also observe a mass-dependent spin transition, with low-mass galaxies exhibiting a discernible alignment signal, at least in high-$q$ regions.

\begin{figure*}[t]
\epsscale{1.15}
\plotone{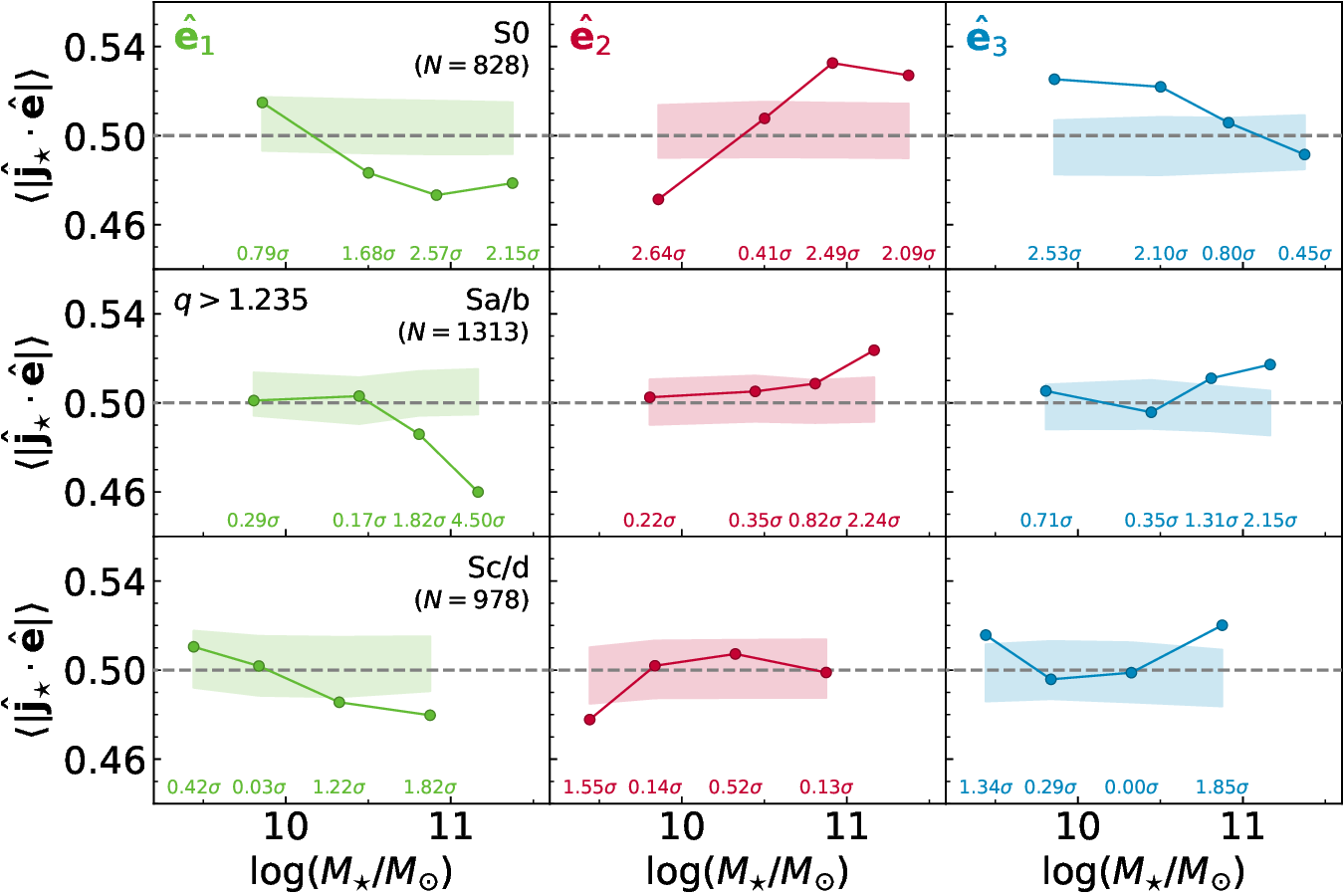}
\caption{Same as Figure \ref{fig:alignall}, but for lenticulars ($-2\leq\mathrm{TType}\leq0$; top), early-type spirals ($1\leq\mathrm{TType}\leq3$; middle), and late-type spirals ($4\leq\mathrm{TType}\leq8$; bottom) in high-$q$ regions. \label{fig:alignmorp}}
\end{figure*}

Figure \ref{fig:alignmorp} presents how the spin--tidal alignment varies with galaxy morphology. To emphasize clear trends, we use the same galaxies in high-$q$ regions as in the bottom panels of Figure \ref{fig:alignall}. Because the stellar and gas spin vectors yield nearly identical results, only stellar spins are shown for simplicity. The top, middle, and bottom panels correspond to lenticulars (S0; $-2 \leq \mathrm{TType} \leq 0$), early-type spirals (Sa/b; $1 \leq \mathrm{TType} \leq 3$), and late-type spirals (Sc/d; $4 \leq \mathrm{TType} \leq 8$), respectively. Overall, the alignment patterns show no statistically significant dependence on morphology, at least among S0 and spiral galaxies. Nevertheless, the results indicate that the S0 and Sa/b populations primarily drive the alignments with $\eone$ and $\etwo$ observed in massive galaxies, while Sc/d galaxies show relatively weak evidence of alignment. With respect to $\ethree$, no significant alignment signal is detected, except for low-mass ($M_* \leq 10^{10.5}\,M_\odot$) S0 galaxies, which tend to align with $\ethree$.

\begin{figure*}[t]
\epsscale{1.15}
\plotone{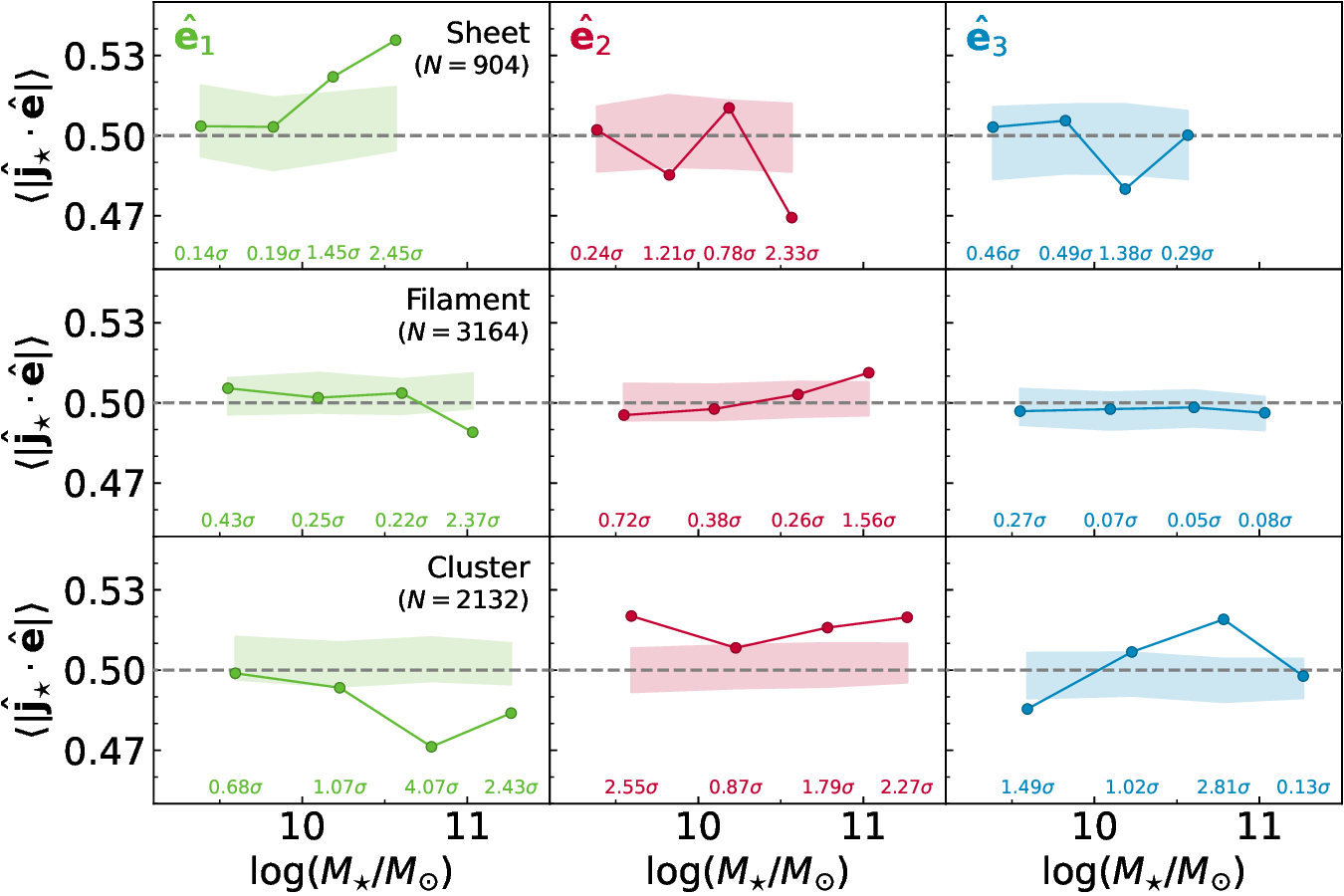}
\caption{Same as Figure \ref{fig:alignall}, but for galaxies residing in the sheet (top), filament (middle), and cluster (bottom) environments. Galaxies in void environments are not shown because of their small sample size. \label{fig:alignweb}}
\end{figure*}

In Figure \ref{fig:alignweb}, we compare the alignments across different large-scale environment types, namely, sheets, filaments, and clusters. The strongest statistical significance is found in clusters, where galaxies tend to align parallel to $\etwo$ over a wide mass range, while massive galaxies ($M_* > 10^{10}\,M_\odot$) show a tendency to align perpendicular to $\eone$. Galaxies in filaments exhibit a broadly similar trend to the total sample in Figure \ref{fig:alignall}. The significance, however, is weaker, and no clear alignment is found with $\ethree$, which corresponds to the direction of the filament spine in filament environments. Interestingly, galaxies in sheet environments show a distinct behavior: their spins preferentially align parallel with $\eone$, the axis perpendicular to the sheet plane, for $M_* \geq 10^{10}\,M_\odot$. 

\begin{figure*}[t]
\epsscale{1.15}
\plotone{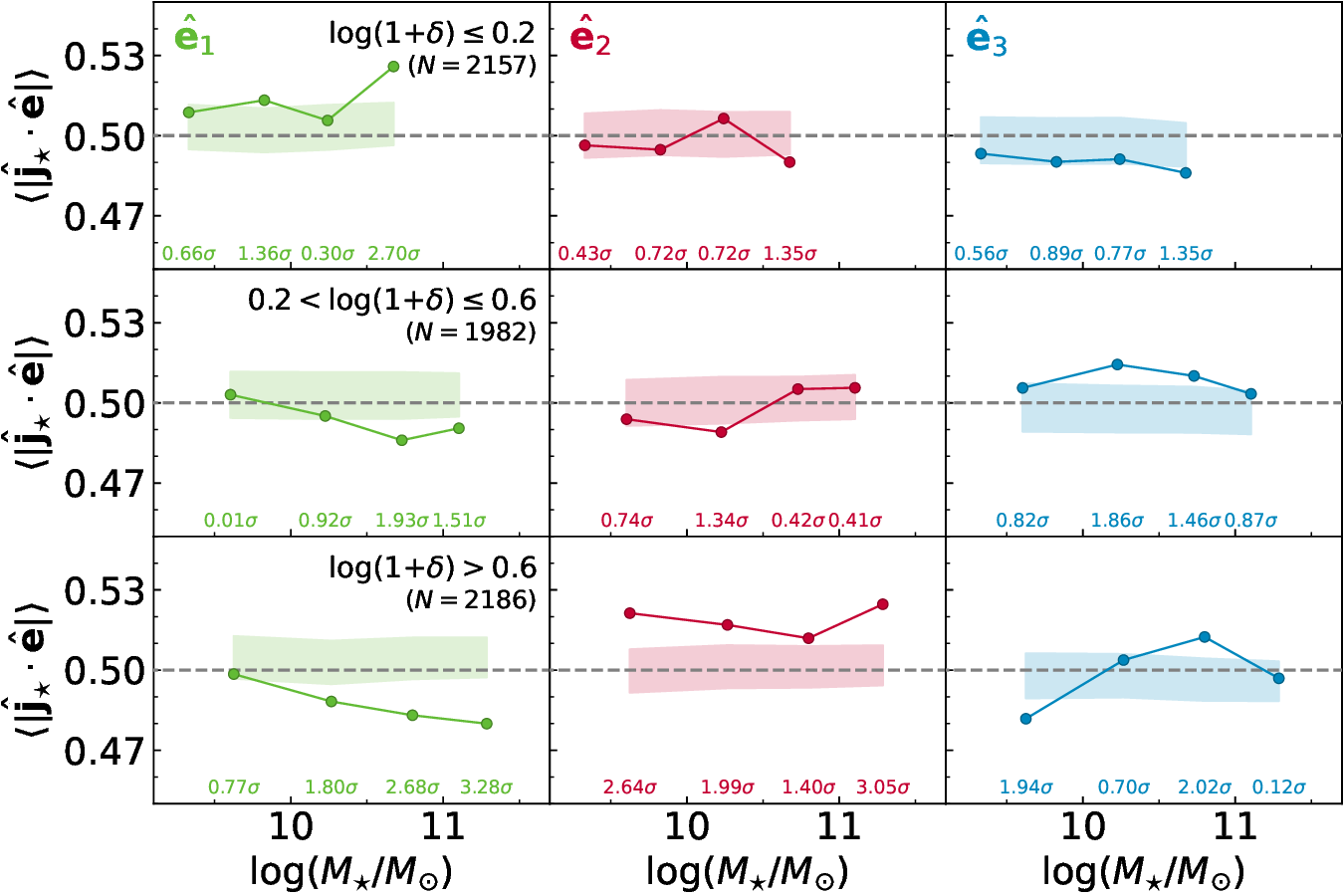}
\caption{Same as Figure \ref{fig:alignall}, but for galaxies residing in high-$\delta$ (top), intermediate-$\delta$ (middle), and low-$\delta$ (bottom) regions. \label{fig:alignden}}
\end{figure*}

\begin{figure*}[ht!]
\epsscale{1.15}
\plotone{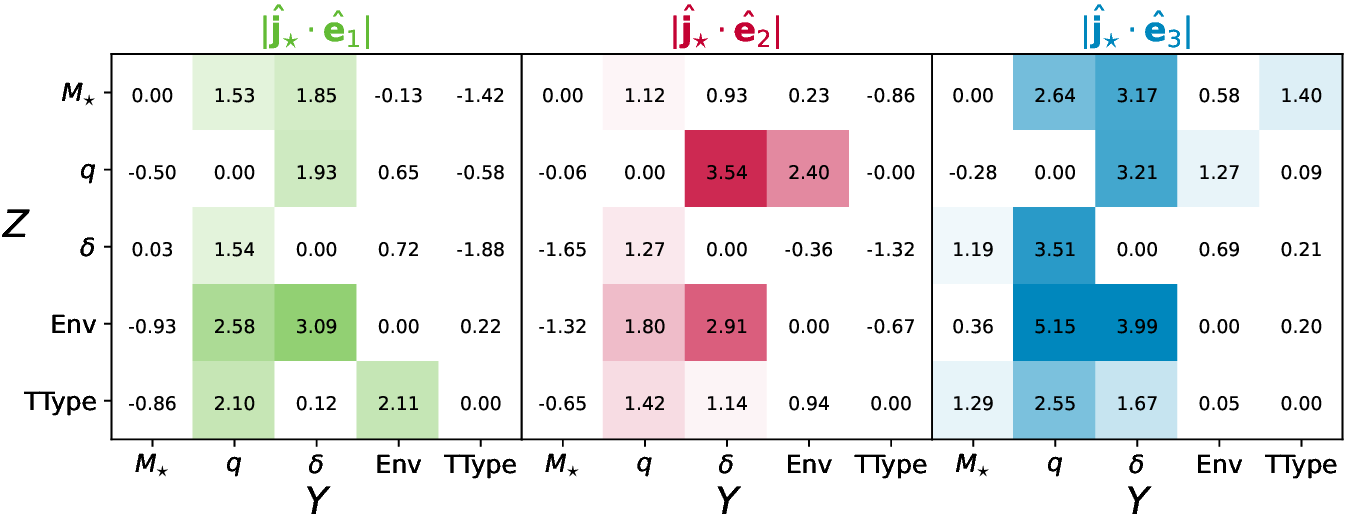}
\caption{Normalized conditional mutual information $I(|\hat{\boldsymbol{j}} \cdot \hat{\boldsymbol{e}}_k|;Y|Z)$ ($k = 1, 2, 3$) for various combinations of $Y$ and $Z$. Each value is normalized by the mean and standard deviation estimated from 1000 randomly shuffled samples. Darker colors indicate stronger importance of $Y$ in determining spin alignments after controlling for $Z$. \label{fig:mi}}
\end{figure*}

In Figure \ref{fig:alignden}, we further examine the dependence of the spin--tidal alignment on environmental overdensity ($\delta \equiv \Delta\rho/\bar{\rho}$). The sample is divided into three subsamples of roughly equal size, corresponding to low-, intermediate-, and high-$\delta$ regions. The alignment trends for these density bins qualitatively resemble those found in sheets, filaments, and clusters, respectively. The mass-dependent trends shown in Figure \ref{fig:alignall} weaken once $\delta$ is divided. Overall, these results demonstrate that the spin--tidal alignment strongly depends on the large-scale environments.

To evaluate the relative importance of galaxy properties in determining spin alignments, we employ the conditional mutual information (CMI; \citealt{1948BSTJ...27..379S, 2017MNRAS.467L...6P}). The CMI, $I(X;Y|Z)$, quantifies how much information $Y$ provides about $X$ beyond what is explained by their mutual correlation with $Z$. When the parameter space spanned by $X$, $Y$, and $Z$ is divided into a total number of $N_X \times N_Y \times N_Z$ bins of small volume, $I(X;Y|Z)$ can be calculated as
\begin{align}
I(X;Y|Z) = &\sum_{a=1}^{N_X} \sum_{b=1}^{N_Y} \sum_{c=1}^{N_Z} \biggl[p(X_a, Y_b, Z_c)\nonumber \\
&\times \log \frac{p(Z_c)\, p(X_a, Y_b, Z_c)}{p(X_a, Z_c)\, p(Y_b, Z_c)}\biggl], 
\end{align}
where $p(X_a,Y_b,Z_c)$ denotes the probability function of the variables $(X,Y,Z)$ falling within the $(a,b,c)$th pixel. A higher CMI indicates that $Y$ retains more significant information about $X$ even after controlling for $Z$. It should be noted that the CMI measures statistical correlations across overall distributions rather than systematic trends in the mean value.

Figure \ref{fig:mi} plots the CMI, $I(X;Y|Z)$, for $X = \anglest$ ($k = 1, 2, 3$), with $Y$ and $Z$ representing various galaxy properties. The values are normalized by the mean and standard deviation derived from 1000 randomly shuffled samples. The most prominent feature is that $I(X;q|Z)$ and $I(X;\delta|Z)$ exhibit consistently high values, indicating the strongest correlations of spin alignments with environmental factors. The high values of both $I(X;\delta|q)$ and $I(X;q|\delta)$ further imply that these two parameters make independent contributions. In contrast, $I(X;M_*|Z)$ remains small, suggesting that stellar mass becomes less influential once environmental effects are considered, as shown in Figure \ref{fig:alignden}. Morphology and environment types also seem to add little information beyond $M_*$ or $\delta$. A comparison of the three principal axes shows broadly similar patterns, though alignments with $\ethree$ yield slightly larger CMI, likely reflecting more intricate correlations among galaxy properties. 

\section{Summary and Discussion} \label{sec:summary}

We have examined the alignment between the spin vectors and the principal axes of the local tidal field using the complete release of the MaNGA IFS survey. The results reveal clear mass-dependent alignment trends. On average, massive galaxies tend to orient their spins perpendicular to the major axis ($\eone$) and parallel to the intermediate axis ($\etwo$). Low-mass galaxies show the opposite tendency with a transition mass of $M_* \sim 10^{10}-10^{10.5}\,M_\odot$. This alignment trend becomes particularly pronounced in regions of high tidal anisotropy. No statistically significant alignment signal is detected with respect to the minor axis ($\ethree$). Galaxy morphology affects the alignment strength more than the preferred direction, with S0 and Sa/b galaxies driving the signal in the massive regime. The alignment patterns vary systematically across large-scale environments: cluster galaxies exhibit the strongest alignment, filament galaxies follow similar but weaker trends, and sheet galaxies show a distinctive parallel alignment with $\eone$. Mutual information analysis suggests that tidal anisotropy ($q$) and overdensity ($\rho$) play the primary roles in determining spin alignments.

Various physical mechanisms that influence galaxy spins are expected to imprint distinct alignment patterns with respect to the local tidal field. Specifically, primordial tidal torques are predicted to produce spin alignments along $\etwo$, as described by the linear TTT \citep[e.g.,][]{2000ApJ...532L...5L, 2009IJMPD..18..173S}. In the nonlinear regime, the vorticity vectors tend to be oriented perpendicular to $\eone$ \citep[e.g.,][]{2013ApJ...766L..15L, 2023ApJ...945...13M}, leading galaxy spins to align in the same direction as the local vorticity. Galaxy mergers occur more frequently along $\ethree$ \citep[e.g.,][]{2015A&A...576L...5T, 2015MNRAS.450.2727T}, and their orbital angular momenta, being perpendicular to the merger axis, tend to produce spin alignments perpendicular to $\ethree$. In our results, the alignment with $\etwo$ can be naturally linked to the TTT. In particular, massive galaxies in regions of high $q$ and high $\rho$ are likely to experience stronger tidal influences, which can enhance the alignment strength with $\etwo$. 

The alignment with $\eone$, however, is less straightforward to interpret. Since low-mass galaxies are likely to be more strongly affected by nonlinear effects, their observed tendency to align parallel to $\eone$ seems inconsistent with theoretical expectations. This result also contrasts with results from hydrodynamical simulations such as IllustrisTNG, which predict parallel alignment with $\eone$ in the massive regime \citep[see, e.g.,][]{2021ApJ...922....6L, 2023ApJ...945...13M}.

Nevertheless, some observational studies have reported results that cannot be fully explained within this theoretical picture. For instance, \citet{2015ApJ...798...17Z} identified a spin transition with respect to $\eone$, from parallel to perpendicular, as halo mass increases. They further showed that the alignment with $\etwo$ strengthens with increasing mass and becomes more pronounced in cluster environments, both consistent with our findings. Likewise, \citet{2023ApJ...951L..26L} found a spin transition among spiral galaxies on void surfaces, from parallel to perpendicular orientations with respect to the direction of maximum compression (i.e., $\eone$), as stellar mass increases. Such observational deviations may point to the potential influence of baryonic physics, such as anisotropic outflows or heating from stellar and active galactic nucleus feedback \citep[e.g.,][]{2017ApJ...834..169T, 2020MNRAS.492.4268S}, whose roles in spin alignments remain poorly understood.

Most IFS-based studies of spin alignment have focused on the orientation of galaxy spins relative to cosmic filaments, revealing various mass-dependent \citep{2020MNRAS.491.2864W, 2022MNRAS.516.3569B} and morphology-dependent \citep{2021MNRAS.504.4626K} spin transitions. Although a direct comparison is difficult, our results show no significant spin alignment signal for $\ethree$, which is somewhat surprising given that filament spines are generally aligned with $\ethree$ \citep[e.g.,][]{2014MNRAS.437L..11T, 2015MNRAS.453L.108L}. A comparative analysis between spin alignments with filaments and those with tidal fields would be an interesting direction for future work.

Lastly, we caution that the alignment trends reported in this work are derived from a tidal field smoothed on a scale of 2\,$h^{-1}$Mpc. This smoothing scale is comparable to the characteristic size of large-scale structures such as filaments and galaxy clusters, and is therefore a commonly adopted choice \citep[e.g.,][]{2007MNRAS.381...41H, 2007MNRAS.375..489H, 2009ApJ...706..747Z, 2015ApJ...798...17Z, 2017MNRAS.468L.123W}. Too small smoothing scales can break the linear approximation of the tidal field, while too large scales can wash out the spin--tidal correlations. However, previous studies have indicated that the detailed trend can vary with the smoothing scale \citep{2012MNRAS.427.3320C, 2013ApJ...762...72T, 2018MNRAS.473.1562W, 2019ApJ...872...37L, 2021ApJ...922....6L, 2023ApJ...945...13M, 2023ApJ...952...82M}, suggesting that future analyses using different scales will be valuable for understanding the scale dependence of the observed alignments.

This study provides one of the first observational investigations of spin--tidal alignment using IFS survey data, revealing clear mass- and environment-dependent alignment trends. The detection of spin alignment with the tidal field highlights the strong influence of large-scale structures in shaping galaxy angular momentum. Our results challenge the conventional picture, indicating that the role of baryonic processes in regulating spin orientations may require further investigation. Future work combining hydrodynamical simulations that incorporate diverse baryonic physics models can help establish a more complete understanding of galaxy spin evolution. Moreover, next-generation IFS surveys with larger samples will enable more detailed statistical analyses of how stellar spin alignments depend on various galaxy properties.


\begin{acknowledgments}
We thank the anonymous referee for useful comments that improved the manuscript. We also thank Huiyuan Wang for helpful correspondence regarding the GEMA-VAC data.
T.O. acknowledges support from the Taiwan National Science and Technology Council under Grants 
Nos. NSTC 112-2112-M-001-034-,
NSTC 113-2112-M-001-011- and
NSTC 114-2112-M-001-004-, and the Academia Sinica Investigator Project Grant No. AS-IV-114-M03 for the period of 2025--2029. 

Funding for the Sloan Digital Sky Survey IV has been provided by the Alfred P. Sloan Foundation, the U.S. Department of Energy Office of Science, and the Participating Institutions. SDSS acknowledges support and resources from the Center for High-Performance Computing at the University of Utah. The SDSS website is \url{www.sdss.org}.

SDSS-IV is managed by the Astrophysical Research Consortium for the Participating Institutions of the SDSS Collaboration including the Brazilian Participation Group, the Carnegie Institution for Science, Carnegie Mellon University, Center for Astrophysics \textbar\ Harvard \& Smithsonian (CfA), the Chilean Participation Group, the French Participation Group, Instituto de Astrof\'isica de Canarias, The Johns Hopkins University, Kavli Institute for the Physics and Mathematics of the Universe (IPMU) / University of Tokyo, the Korean Participation Group, Lawrence Berkeley National Laboratory, Leibniz Institut f\"ur Astrophysik Potsdam (AIP), Max-Planck-Institut f\"ur Astronomie (MPIA Heidelberg), Max-Planck-Institut f\"ur Astrophysik (MPA Garching), Max-Planck-Institut f\"ur Extraterrestrische Physik (MPE), National Astronomical Observatories of China, New Mexico State University, New York University, University of Notre Dame, Observat\'orio Nacional / MCTI, The Ohio State University, Pennsylvania State University, Shanghai Astronomical Observatory, United Kingdom Participation Group, Universidad Nacional Aut\'onoma de M\'exico, University of Arizona, University of Colorado Boulder, University of Oxford, University of Portsmouth, University of Utah, University of Virginia, University of Washington, University of Wisconsin, Vanderbilt University, and Yale University. \vspace{0.0pt}

\end{acknowledgments}





%
\facilities{Sloan(MaNGA)}\vspace{0.0pt}

\software{Astropy \citep{2013A&A...558A..33A, 2018AJ....156..123A, 2022ApJ...935..167A}, Marvin \citep{2019AJ....158...74C}, Matplotlib \citep{2007CSE.....9...90H}, NumPy \citep{2020Natur.585..357H}, PaFit \citep{2006MNRAS.366..787K}}


\appendix 
\section{Gas Spin Alignments} \label{append:gas}

\begin{figure*}[t]
\epsscale{1.13}
\plotone{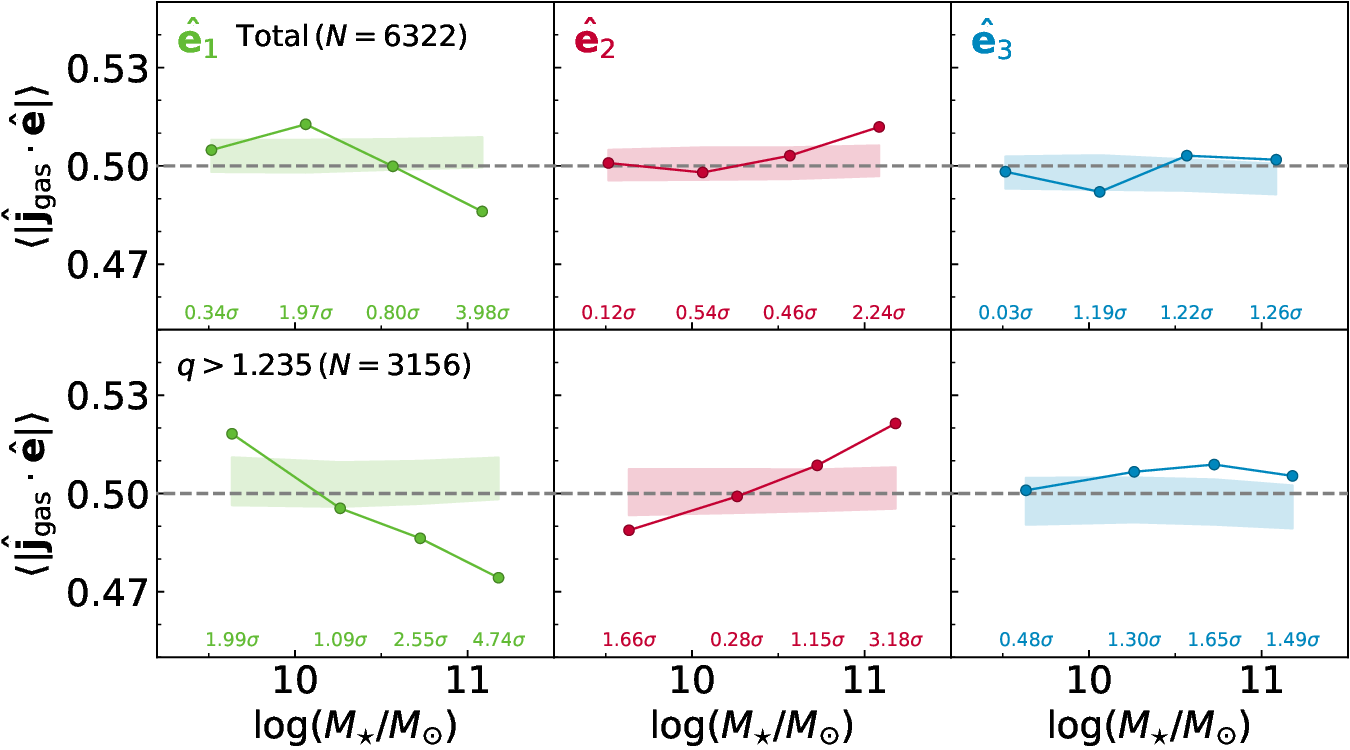}
\caption{Same as Figure \ref{fig:alignall}, but showing the spin--tidal alignments for the gas spin vectors. \label{fig:aligngas}}
\end{figure*}

Figure \ref{fig:aligngas} shows the spin--tidal alignment for gas spin vectors. The top and bottom panels correspond to the total sample and the subsample residing in regions of strong tidal anisotropy ($q > q_\mathrm{med} = 1.235$), respectively. The trends closely follow those found for the stellar components discussed in Section \ref{sec:align}. Overall, the gas spins exhibit the same qualitative mass- and environment-dependent behavior as the stellar spins, suggesting that both components trace the large-scale tidal field in a consistent manner.

\vspace{8pt}
\section{Twofold Ambiguity of Spin Vectors} \label{append:sign}

\begin{figure*}[t]
\epsscale{1.13}
\plotone{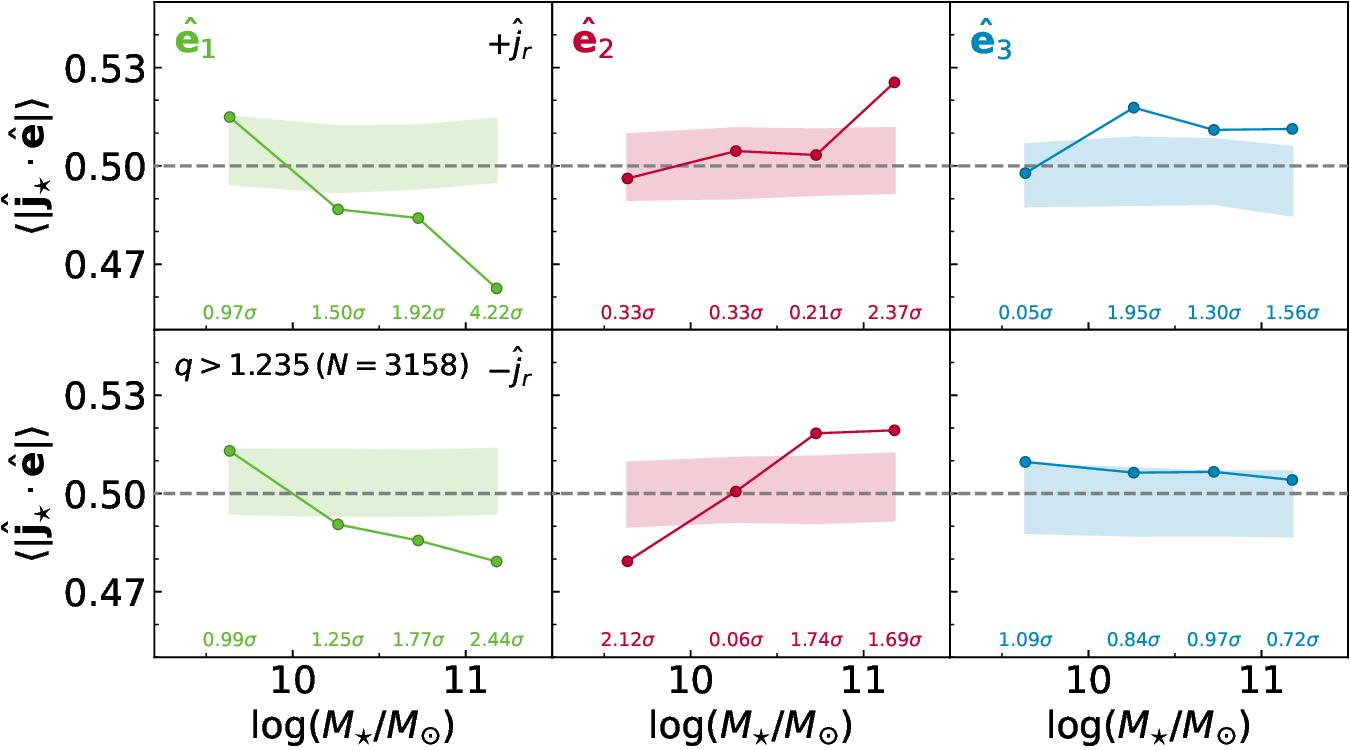}
\caption{Same as the bottom panels of Figure \ref{fig:alignall}, but recomputed with the sign of $\hat{j}_r$ fixed to either $+$ (top) or $-$ (bottom). \label{fig:ambiguity}}
\end{figure*}

As discussed in Section \ref{subsec:manga}, the twofold ambiguity in the sign of $\hat{j}_r$ leads to two possible orientations of a galaxy's 3D spin vector. Figure \ref{fig:ambiguity} tests the impact of this ambiguity by recomputing the spin--tidal alignment in high-$q$ regions (same as the bottom panels of Figure \ref{fig:alignall}) with all signs of $\hat{j}_r$ fixed to either $+$ (radially outward) or $-$ (radially inward). Both cases reproduce the same qualitative alignment trends as the fiducial analysis, showing only a modest reduction in the detection significance.

Galaxies whose inferred 3D spin orientation changes substantially under different sign choices tend to be less reliable. This consideration motivates our use of the averaged orientation, which provides a more stable ensemble statistic. For this reason, we adopt the averaged orientation as our fiducial choice, while noting that our main conclusions are qualitatively insensitive to the specific sign convention.


\newpage\
\newpage
\bibliography{main}{}
\bibliographystyle{aasjournalv7}



\end{document}